\documentclass[12pt]{iopart}

\usepackage{iopams}
\usepackage{dsfont}

\usepackage{graphicx,graphics,color}
\usepackage{colortbl}

\usepackage{hyperref}

\begin{document}

\newcommand{\nn}{\nonumber}
\newcommand{\bra}{\langle}
\newcommand{\ket}{\rangle}
\newcommand{\del}{\partial}
\newcommand{\vt}{\vec}
\newcommand{\dg}{^{\dag}}
\newcommand{\cg}{^{*}}
\newcommand{\vep}{\varepsilon}
\newcommand{\suml}{\sum\limits}
\newcommand{\prodl}{\prod\limits}
\newcommand{\intl}{\int\limits}
\newcommand{\til}{\widetilde}
\newcommand{\mcl}{\mathcal}
\newcommand{\mfk}{\mathfrak}
\newcommand{\ds}{\displaystyle}
\newcommand{\mod}{\mathrm{\,mod\,}}
\renewcommand{\dim}{\mathrm{dim}}
\renewcommand{\Re}{\mathrm{Re}}
\renewcommand{\Im}{\mathrm{Im}}
\renewcommand{\b}{\overline}

\eqnobysec

\title[Bose-Einstein condensate in a triple-well trap]
{Dynamics of a Bose-Einstein condensate in a symmetric triple-well trap}

\author{Thiago F Viscondi and K Furuya}

\address{Instituto de Física ``Gleb Wataghin'', Universidade Estadual de Campinas - UNICAMP, 13083-859, Campinas, SP, Brazil}

\ead{viscondi@ifi.unicamp.br}

\begin{abstract}
We present a complete analysis of the dynamics of a Bose-Einstein condensate trapped in a symmetric triple-well potential. Our
classical analogue treatment, based on a time-dependent variational method using $\mathrm{SU}(3)$ coherent states, includes the
parameter dependence analysis of the equilibrium points and their local stability, which is closely related to the condensate
collective behaviour. We also consider the effects of off-site interactions, and how these `cross-collisions' may become relevant
for a large number of trapped bosons. Besides, we have shown analytically, by means of a simple basis transformation in the
single-particle space, that an integrable sub-regime, known as twin-condensate dynamics, corresponds in the classical phase space
to invariant surfaces isomorphic to the unit sphere. However, the quantum dynamics preserves the twin-condensate defining
characteristics only partially, thus breaking the invariance of the associated quantum subspace. Moreover, the periodic geometry
of the trapping potential allowed us to investigate the dynamics of finite angular momentum collective excitations, which can be
suppressed by the emergence of chaos. Finally, using the generalized purity associated to the $\mathrm{su}(3)$ algebra, we were
able to quantify the dynamical classicality of a quantum evolved system, as compared to the corresponding classical trajectory.
\end{abstract}


\section{Introduction}
\label{intro}

The experimental observation of the Bose-Einstein condensation in systems of ultracold dilute alkali atomic clouds
\cite{becreview} has led to an active development of various studies concerning collective quantum phenomena. The special
property of coherence presented in condensates of a large number of atoms allowed mean-field and semiclassical approaches to
discover the fascinating dynamics of such a large quantum system.

The prospect of using magneto-optical traps with diverse confining geometries allowed the study of various quantum dynamical
regimes, also including the possibility of testing the frontier between the quantum and classical mechanics through the increase
of the total particle number to the macroscopic scale.

Several surveys have been performed in the case of a Bose-Einstein condensate trapped in a double well-potential under the
two-mode approximation. Various subjects were considered in these studies, such as tunneling dynamics \cite{tunnel}, entanglement
of modes \cite{entanglement} and quantum phase transition \cite{qpt2well}. More recently the effects of the so-called
\textit{cross-collisions} have also been considered, and it was shown that these usually neglected higher order approximation
terms can become quite important for large numbers of trapped bosons \cite{crosscoll}.

The dynamics of condensates in a triple-well potential has already been investigated a few times in the literature \cite{3well}.
However, little attention has been paid to the specific case in which all three wells are identically coupled in pairs
\cite{3wellring}. This arrangement is very interesting because it represents the simplest trapping potential where we can observe
the rotation of the condensate.

In the present work we examine the dynamics of a condensate in a symmetric triple-well potential, under the three-mode
approximation and considering the cross-collision terms. First, we investigate the classical analogue of the model and its phase
space structure, by determining its fixed points and the Hamiltonian flow in the vicinity of these special solutions, which is
closely related to the condensate collective excitations. Then, we choose the coherent states located in the proximity of the
classical stationary states as initial quantum states. This enables us to capture in the \textit{quantum dynamics} the
characteristics associated to each of the classical fixed points. This strategy has been used before to observe the effect of
chaos in the entanglement dynamics \cite{chaosvsregular}.

In section \ref{model} we derive the model in the three-mode approximation, including the cross-collisional terms, and the
corresponding Hamiltonian in terms of the $\mathrm{SU}(3)$ group generators. We also present the classical approximation based on
the \textit{time-dependent variational principle} (TDVP) \cite{classical} and the $\mathrm{SU}(3)$ coherent states
\cite{coherent}. The equilibrium points of the classical analogue equations of motion are divided into three classes:
\textit{twin-condensate states}, \textit{single depleted well states} and \textit{vortex states}. Here we show in detail the
stability diagrams of the fixed points as functions of two collision parameters. The behaviour of the lowest energy fixed point
and the occurrence of a quantum phase transition has been reported recently \cite{qpt3well}.

In section \ref{dynamics} we relate a simple permutation symmetry of the Hamiltonian to a transformation of the single-particle
basis that leads to the twin-condensate subspace. Then, we demonstrate that the classically invariant subspace of
twin-condensates is not fully preserved quantum mechanically. In addition, we compare the quantum population dynamics in the
local modes with the corresponding classical evolution. Particularly interesting is the behaviour of the vortex states, for which
we show the dynamics of a quantity associated with the condensate angular momentum along the symmetry axis of the trapping
potential. The variation of the collision parameters can lead the vortex solutions to chaotic dynamics, and consequently
inhibiting the condensate rotation in a preferred sense. It is notable that this model present chaotic behaviour at the classical
level. Moreover, similarly to previous results for the double-well condensate model \cite{qpt2well} with the measure known as
\textit{generalized purity} \cite{purity}, we make a quantitative comparison between the quantum evolution and the classical
approximation based on the coherent states. Section \ref{conclusion} is reserved to our conclusions and final considerations.

\section{Model}
\label{model}

Suppose a spinless boson of mass $m$ trapped in a potential $V(\vt{r}) $, whose equivalent minima are located at
$\vt{r}_{j}=(x_{j},y_{j},0)$, $j=1,2,3$:

\begin{equation}
  V(\vt{r})=\frac{m\omega^{2}}{18q_{0}^{4}}\prod_{j=1}^{3}\left[(x-x_{j})^{2}+(y-y_{j})^{2}\right]
  +\frac{m\omega^{2}}{2}z^{2}. \label{eq1}
\end{equation}

Note that the potential is harmonic in $z$-direction with angular frequency $\omega$ and we choose $V(\vt{r}_{j})=0$. The
parameter $q_{0}$ represents the distance between the potential minima and the origin of the coordinate system. For simplicity,
we also choose $V(\vt{r})$ so that the harmonic approximation $V_{j}^{(2)}(\vt{r})$ around each minimum is isotropic with angular
frequency $\omega$. We can assign to each potential $V_{j}^{(2)}(\vt{r})$, taken independently, a localized ground state:

\begin{equation}
  \bra\vt{r}|u_{j}\ket=u_{j}(\vt{r})=\frac{1}{\pi^{\frac{3}{4}}d^{\frac{3}{2}}}\exp\left[
  -\frac{(x-x_{j})^{2}+(y-y_{j})^{2}+z^2}{2d^{2}}\right].
  \label{eq2}
\end{equation}

The Gaussian  width is described by the parameter $d=\sqrt{\hbar/m\omega}$. Considering that $q_{0}\gg d$, the states
$|u_{j}\ket$ are almost orthogonal, since $\vep=\bra u_{j}|u_{k}\ket=\exp\left[-3q_{0}^{2}/4d^{2}\right]\ll1$, for $j\neq k$.

The states $|u_{j}\ket$ are not eigenstates of the full single-particle Hamiltonian $H=\vt{p}^{2}/2m+V(\vt{r})$, but these
localized states generate the same eigenspace of the three lowest energy states of $H$ at first order in $\vep$. That is, we have
the following single-particle lowest energy states at $O(\vep)$:

\begin{equation}
  \left\{\eqalign{
  |e_{1}\ket=\frac{1}{\sqrt{3}}\left(|u_{1}\ket+|u_{2}\ket+|u_{3}\ket\right)\cr
  |e_{2}\ket=\frac{1}{\sqrt{6}}\left(|u_{1}\ket+|u_{2}\ket-2|u_{3}\ket\right)\cr
  |e_{3}\ket=\frac{1}{\sqrt{2}}\left(|u_{1}\ket-|u_{2}\ket\right)}\right.
  \label{eq3}
\end{equation}

The states $|e_{2}\ket$ and $|e_{3}\ket$ are degenerated and the energy of these states differs from the energy of $|e_{1}\ket$
by $3\Omega$, with the tunneling rate defined by:

\begin{equation}
  \Omega\equiv\bra u_{j}|V(\vt{r})-V_{j}^{(2)}(\vt{r})|u_{k}\ket+\vep\frac{3\hbar\omega}{2};\qquad j\neq k.
  \label{eq4}
\end{equation}

Generally, for $q_{0}\gg d$, $\Omega$ is negative and $|e_{1}\ket$ is the single-particle ground state, whose energy difference
for the next two states is proportional to $\vep$.

So far we studied only the single-particle potential. Now, considering a system of $N$ condensate bosons, we introduce the
many-particle Hamiltonian:

\begin{equation}
  \hat{H}=\int\rmd^{3}r\,\hat{\psi}\dg(\vt{r})H\hat{\psi}(\vt{r})
  +\frac{1}{2}\int\rmd^{3}r\,\rmd^{3}r'\hat{\psi}\dg(\vt{r})\hat{\psi}\dg(\vt{r}')
  \mathcal{V}(\vt{r},\vt{r}')\hat{\psi}(\vt{r}')\hat{\psi}(\vt{r}).
  \label{eq5}
\end{equation}

\noindent where $H$ is the single-particle Hamiltonian and $\mathcal{V}(\vt{r},\vt{r}')$ is the pair interaction potential. Only
pair interactions are significant, because we assume dilute condensate gas, characterized by a high mean free path, which makes
higher order collisions unlikely. For particles of momentum approaching zero, as in the case of a condensate, a good
approximation for the interaction potential is the following effective potential \cite{pseudpot}:

\begin{equation}
  \mathcal{V}(\vt{r},\vt{r}')=\frac{4\pi\hbar^{2}a}{m}\delta(\vt{r}-\vt{r}')=V_{0}\delta(\vt{r}-\vt{r}').
  \label{eq6}
\end{equation}

In the previous equation we defined the $s$-wave scattering length, denoted by $a$. Now, supposing that the states of \eref{eq3}
are the only significantly populated, we get our second approximation in \eref{eq5}, namely, the \textit{three-mode
approximation}:

\begin{equation}
  \hat{\psi}(\vt{r})\approx\bra\vt{r}|u_{1}\ket a_{1}+\bra\vt{r}|u_{2}\ket a_{2}+\bra\vt{r}|u_{3}\ket a_{3};
  \label{eq7}
\end{equation}

\noindent where we defined $a_{j}$ as the annihilation operator for the state $|u_{j}\ket$. Therefore, the field operator
$\hat{\psi}(\vt{r})$ may be expanded in the local modes described in \eref{eq2}, which generate the same single-particle space as
the energy states of \eref{eq3}. Substituting \eref{eq6} and \eref{eq7} in \eref{eq5}, we obtain the following
Hamiltonian\footnote{Note that we use the bosonic canonical commutation relations to simplify the final expression for $\hat{H}$.
We also discard the terms dependent only on the conserved total number of particles, as such terms do not change the system
dynamics for a fixed value of $N$.}, retaining only the terms up to order $\vep^{\frac{3}{2}}$:

\begin{equation}
  \hat{H}=\Omega'\suml_{j\neq k}a\dg_{j}a_{k}+\kappa\suml_{j}a^{\dagger 2}_{j}a^{2}_{j}
  -2\Lambda\widetilde{\suml_{j,k,m}}a\dg_{j}a_{j}a\dg_{k}a_{m}.
  \label{eq8}
\end{equation}

The third summation symbol in \eref{eq8} indicates the sum over three different indices. In the previous equation we defined the
effective tunneling rate $\Omega'=\Omega+2\Lambda(N-1)$ and the following collision parameters:

\begin{equation}
  \eqalign{
  \kappa=\frac{V_{0}}{2}\int\rmd^{3}r\, u_{j}^{4}(\vt{r})=\frac{V_{0}}{2^{\frac{5}{2}}\pi^{\frac{3}{2}}d^{3}};\cr
  \Lambda=\frac{V_{0}}{2}\int\rmd^{3}r\, u_{j}^{3}(\vt{r})u_{k}(\vt{r})=\kappa\vep^{\frac{3}{2}};\qquad j\neq k.}
  \label{eq9}
\end{equation}

The \textit{self-collision rate} $\kappa$ is proportional to the collision frequency between bosons in the same potential well,
while the \textit{cross-collision rate} $\Lambda$ is proportional to the collision frequency between bosons coming from different
local modes. Note that the effective tunneling rate $\Omega'$ depends on the cross-collision parameter and the total number of
trapped bosons. Also, observe that we have $|\Lambda|\ll|\kappa|$ for $\vep\ll1$. Although the cross-collision rate has a lower
order of magnitude than $\Omega$, the product $N\Lambda$ in $\Omega'$ may be relevant for large numbers of trapped bosons.
Therefore, the cross-collisions become important in comparing theoretical and experimental results, because they include
significant effects of the number of particles in the effective tunneling rate.

Since $\hat{H}$ conserves the total number of trapped particles under unitary evolution, we can use the following homomorphism
between the bilinear bosonic operators and the generators of the $\mathrm{SU}(3)$ group:

\begin{equation}
  \eqalign{
  Q_{1}=\frac{1}{2}(a\dg_{1}a_{1}-a\dg_{2}a_{2}),\qquad& Q_{2}=\frac{1}{3}(a\dg_{1}a_{1}+a\dg_{2}a_{2}-2a\dg_{3}a_{3}),\cr
  J_{k}=i(a\dg_{k}a_{j}-a\dg_{j}a_{k}),\qquad& P_{k}=a\dg_{k}a_{j}+a\dg_{j}a_{k};}
  \label{eq10}
\end{equation}

\noindent for $k=1,2,3$ and $j=(k+1)\mod3+1$. The Hamiltonian \eref{eq8} can be rewritten in terms of the operators \eref{eq10},
characterizing $\mathrm{SU}(3)$ as the dynamical group of the system\footnote{Here again we discard some terms dependent only on
the fixed total number of particles N.}:

\begin{equation}
  \eqalign{
  \hat{H}=&\left(\Omega'-2\Lambda\frac{N}{3}\right)(P_{1}+P_{2}+P_{3})
  +\frac{\kappa}{2}(4Q_{1}^{2}+3Q_{2}^{2})\cr
  &+\Lambda[2Q_{1}(P_{1}-P_{3})+Q_{2}(2P_{2}-P_{1}-P_{3})].}
  \label{eq11}
\end{equation}

Notice that the tunneling term is linear in the $\mathrm{SU}(3)$ generators, while the collision terms are quadratic. Within the
three-mode approximation, a single-particle operator $A$ and its corresponding many-particle operator $\hat{A}$ are related by:

\begin{equation}
  \hat{A}=\suml_{j,k=1}^{3}\bra u_{j}|A|u_{k}\ket a\dg_{j}a_{k}.
  \label{eq12}
\end{equation}

Therefore, considering terms up to first order in $\vep$, we can find relations between the generators of $\mathrm{SU}(3)$ and
basic system observables. First, we have the following condensate position operators, obtained by choosing
$\vt{r}_{1}=\left(-\frac{1}{2}q_{0},\frac{\sqrt{3}}{2}q_{0},0\right)$,
$\vt{r}_{2}=\left(-\frac{1}{2}q_{0},-\frac{\sqrt{3}}{2}q_{0},0\right)$ and $\vt{r}_{3}=\left(q_{0},0,0\right)$:

\begin{equation}
  \eqalign{
  \hat{x}=-\frac{3q_{0}}{2}Q_{2}+\frac{\vep q_{0}}{2}\left[\frac{1}{2}(P_{1}+P_{3})-P_{2}\right];\cr
  \hat{y}=\sqrt{3}q_{0}Q_{1}+\frac{\sqrt{3}\vep q_{0}}{4}(P_{1}-P_{3}).}
  \label{eq13}
\end{equation}

The operators $J_{k}$, $k=1,2,3$, generate a subalgebra of $\mathrm{su}(3)$ isomorphic to $\mathrm{su}(2)$, the angular momentum
algebra. Nevertheless, linear combinations of these operators are directly proportional to the \textit{linear momentum} operators
of the condensate:

\begin{equation}
  \eqalign{
  \hat{p}_{x}=\frac{3\vep q_{0}}{4d^{2}}\left(J_{3}-J_{1}\right);\\
  \hat{p}_{y}=\frac{\sqrt{3}\vep q_{0}}{4d^{2}}\left(J_{1}+J_{3}-2J_{2}\right).}
  \label{eq14}
\end{equation}

However, the  condensate \textit{angular momentum} operator along the z-axis is proportional to the sum
$J_{S}=J_{1}+J_{2}+J_{3}$:

\begin{equation}
  \hat{L}_{z}=\frac{\sqrt{3}\vep q_{0}^{2}}{4d^{2}}J_{S}.
  \label{eq15}
\end{equation}

\subsection{Classical Approximation}

Mean field theories and classical trajectories are used widely in the literature for the treatment of condensate dynamics. Here
we introduce our classical approach to the problem based on the \textit{time-dependent variational principle}. Considering
independent variations in an arbitrary state $|\psi\ket$ of Hilbert space and its conjugate $\bra\psi|$, it is known that the
extremization of the following action functional is equivalent to the Schr\"{o}dinger equation\footnote{In what follows we make
$\hbar=1$.}:

\begin{equation}
  S=\intl_{t_{i}}^{t_{f}}\rmd t\,\left[\frac{i}{2}\frac{\bra\psi|\dot{\psi}\ket-\bra\dot{\psi}|\psi\ket}{\bra\psi|\psi\ket}
  -\frac{\bra\psi|\hat{H}|\psi\ket}{\bra\psi|\psi\ket}\right].
  \label{eq16}
\end{equation}

We can make approximations to the Schr\"{o}dinger equation by a suitable parametrization of the time dependence of the state
$|\psi\ket$, restricting the state evolution to a subspace of the Hilbert space. The classical approximation is achieved when the
state is restricted to the nonlinear subspace constituted only by the coherent states associated with the dynamical group
\cite{classical}. For the $\mathrm{SU}(3)$ bosonic representations, we have the following coherent states \cite{coherent}:

\begin{equation}
  \eqalign{
  |N;w_{1},w_{2}\ket&=\suml_{n_{1}+n_{2}+n_{3}=N}\left(\frac{N!}{n_{1}!n_{2}!n_{3}!}\right)^{\frac{1}{2}}
  \frac{w_{1}^{n_{1}}w_{2}^{n_{2}}|n_{1},n_{2},n_{3}\ket}{(|w_{1}|^{2}+|w_{2}|^{2}+1)^{\frac{N}{2}}}\cr
  &=\frac{1}{\sqrt{N!}}\left[\frac{w_{1}a\dg_{1}+w_{2}a\dg_{2}+a\dg_{3}}{\sqrt{|w_{1}|^{2}+|w_{2}|^{2}+1}}\right]^{N}|0\ket.}
  \label{eq17}
\end{equation}

The variables $w_{1},w_{2}\in\mathds{C}$ parametrize the subspace composed of coherent states for a fixed number particles $N$.
The states $\{| n_{1},n_{2},n_{3}\ket\}$ form the basis of the three-mode bosonic Fock space, where $n_{j}\in\mathds{N}$ is the
occupation number associated with the local state $|u_{j}\ket$. Applying the variational principle to the action functional
\eref{eq16} with $|\psi\ket=|N;w_{1},w_{2}\ket$, we obtain the classical equations of motion for the complex variables $w_{j}$:

\begin{equation}
  \rmi\dot{w}_{j}=\frac{|w_{1}|^{2}+|w_{2}|^{2}+1}{N}
  \left[(|w_{j}|^{2}+1)\frac{\del\mathcal{H}}{\del w\cg_{j}}+w_{j}w\cg_{k}\frac{\del\mathcal{H}}{\del w\cg_{k}}\right];
  \label{eq18}
\end{equation}

\noindent for $j,k=1,2$ and $k\neq j$. The quantity $\mathcal{H}(w\cg_{1},w\cg_{2},w_{1},w_{2})$ is the effective classical
Hamiltonian, given by the average of $\hat{H}$ in the coherent states:

\begin{equation}
  \fl\eqalign{
  \frac{\mathcal{H}}{N}&=\frac{\bra N;w_{1},w_{2}|\hat{H}|N;w_{1},w_{2}\ket}{N}\cr
  &=\Omega\left\{(1+2\mu)\frac{w_{1}^{*}w_{2}+w_{2}^{*}w_{1}+w_{1}+w_{1}^{*}+w_{2}+w_{2}^{*}}{1+|w_{1}|^{2}+|w_{2}|^{2}}
  +\chi\frac{\left(|w_{1}|^{4}+|w_{2}|^{4}+1\right)}{(|w_{1}|^{2}+|w_{2}|^{2}+1)^{2}}\right.\cr
  &\phantom{=}\left.-2\mu\frac{\left[|w_{1}|^{2}(w_{2}^{*}+w_{2})+|w_{2}|^{2}(w_{1}^{*}+w_{1})+w^{*}_{1}w_{2}+w_{2}^{*}w_{1}\right]}
  {(|w_{1}|^{2}+|w_{2}|^{2}+1)^{2}}\right\}.}
  \label{eq19}
\end{equation}

In the previous equation we introduced new collision parameters, which allowed us to eliminate the dependence on $N$ from the
average energy per particle and also prevented us from problems of divergence in the macroscopic limit $N\rightarrow\infty$. The
quantities $\chi$ and $\mu$ as functions of the old collision parameters are given by:

\begin{equation}
  \chi=\frac{\kappa(N-1)}{\Omega},\qquad\mu=\frac{\Lambda(N-1)}{\Omega}.
  \label{eq20}
\end{equation}

Note that $|\chi|\gg|\mu|$, because supposedly we have $|\kappa|\gg|\Lambda|$. Substituting the Hamiltonian \eref{eq19} into
\eref{eq18}, we obtain the classical equations of motion for the three-mode condensate:

\begin{equation}
  \fl\eqalign{
  \rmi\dot{w}_{j}=&
  \Omega\left\{(1+2\mu)(1-w_{j})(w_{j}+w_{k}+1)+2\chi\frac{w_{j}(|w_{j}|^{2}-1)}{|w_{j}|^{2}+|w_{k}|^{2}+1}\right.\cr
  &\left.-2\mu\frac{[(1-w_{j})w\cg_{k}(w_{j}+w_{k}+w_{j}w_{k})-(1+w_{j})w_{k}(|w_{j}|^{2}-1)]}{|w_{j}|^{2}+|w_{k}|^{2}+1}\right\};}
  \label{eq21}
\end{equation}

\noindent again for $j,k=1,2$ and $k\neq j$. The TDVP based classical approximation with coherent states is exact for any linear
Hamiltonian in the dynamical group generators, i.e., the classical equations of motion \eref{eq21} reproduce the correct quantum
evolution of a coherent state in the case of non-interacting bosons, for $\chi=\mu=0$. The approximation also becomes exact in
the classical limit, which coincides with the macroscopic limit $N\rightarrow\infty$, since the total number of trapped bosons
plays a role equivalent to the reciprocal of $\hbar$ \cite{macrolimit}. However, considering $\chi$ and $\mu$ as independent
parameters, we see that the equations of motion \eref{eq21} and the Hamiltonian per particle \eref{eq19} are independent of $N$,
and therefore represent the \textit{exact results} in the classical-macroscopic limit.

The equations of motion \eref{eq18} assume the canonical form under the following change of dynamical variables:

\begin{equation}
  w_{j}=\left(\frac{K_{j}}{N-K_{1}-K_{2}}\right)^{\frac{1}{2}}\rme^{-\rmi\phi_{j}};
  \label{eq22}
\end{equation}

\noindent for $j=1,2$. The canonical variables $K_{j}$ and $\phi_{j}$ have an useful and immediate physical interpretation. The
variable $K_{j}$ is the mean bosonic occupation in the $j$-th potential well, and $(N-K_{1}-K_{2})$ is the mean population in the
third well due to the conservation of the total number of particles. The angular variable $\phi_{j}$ represents the phase
difference between the portions of the condensate located at the $j$-th and third wells.

\subsection{Equilibrium Points}

\subsubsection{Twin-condensates}

According to the equations of motion \eref{eq21}, the classical phase space of the condensate has three invariant subspaces
arising from the discrete rotational symmetry of the trapping potential, which are described by the following conditions:

\begin{equation}
  w_{1}=w_{2},\qquad w_{1}=1,\qquad w_{2}=1.
  \label{eq24}
\end{equation}

Each one of these equivalent invariant subspaces follows from the equality in phase and mean occupation of a pair of local modes,
according to the transformation \eref{eq22}. Therefore, if two localized condensates have the same initial classical state, so
they remain identical during their classical evolution. This integrable sub-regime of the model is known as
\textit{twin-condensate dynamics}.

The majority of the equilibrium points of the equations of motion \eref{eq21} are located in the twin-condensate subspaces under
the additional constraint $w_{1},w_{2}\in\mathds{R}$. Therefore, by making $w_{1}=w_{2}\in\mathds{R}$ and $\dot{w_{j}}=0$ in
\eref{eq21}, we obtain the following polynomial equation for the location of the twin-condensate fixed points:

\begin{equation}
  4(1+\mu)w_{1}^{4}-2(1+\chi+4\mu)w_{1}^{3}+6\mu w_{1}^{2}+(2\chi-1)w_{1}-1-2\mu=0.
  \label{eq25}
\end{equation}

The above equation clearly has the solution $w_{1}=w_{2}=1$, regardless of the choice of the collision parameter values. This
equilibrium point, according to the interpretation of the transformation \eref{eq22}, corresponds to the state with identical
phase and mean occupation in all three local modes. We denote this solution by $s_{1}$, which is the intersection of the three
invariant surfaces in \eref{eq24}. Excluding the solution $s_{1}$ from \eref{eq25}, we obtain a cubic equation for the remaining
real fixed points:

\begin{equation}
  w_{1}^{3}+\frac{(1-\chi-2\mu)}{2(1+\mu)}w_{1}^{2}+\frac{(1-\chi+\mu)}{2(1+\mu)}w_{1}+\frac{(1+2\mu)}{4(1+\mu)}=0.
  \label{eq26}
\end{equation}

The number of real solutions of the equation \eref{eq26} depends on the sign of the following discriminant:

\begin{equation}
  \fl\eqalign{
  \Delta=&-\chi^{4}-2(3+7\mu)\chi^{3}+3(2-11\mu^{2})\chi^{2}+2(5+12\mu-18\mu^{2}-52\mu^{3})\chi\cr
  &+2(9+76\mu+228\mu^{2}+264\mu^{3}+76\mu^{4}).}
  \label{eq27}
\end{equation}

\begin{figure}[htbp]
  \centering
  \includegraphics[width=0.6\textwidth]{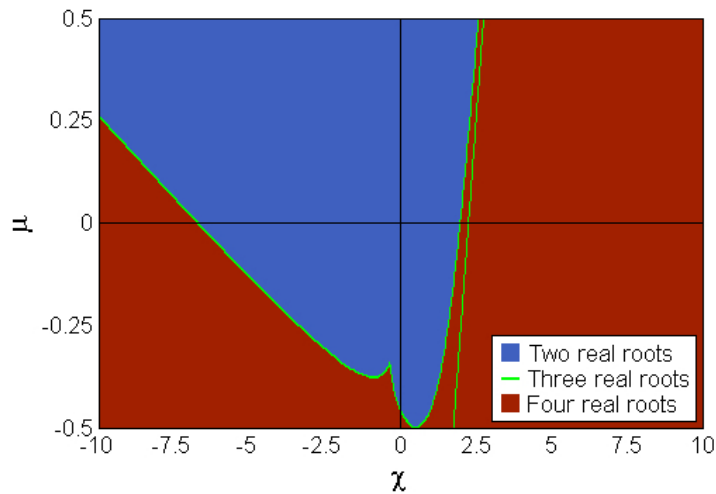}
  \caption{[colour] Number of real solutions of equation \eref{eq25} as a function of the collision parameters. The light
  grey [green] curves represent the transitional situation of existence of only three real solutions. Note that only the quadrants where
  $\chi\mu>0$ are physically accessible, since the two collision parameters must have the same sign.}
  \label{fig1}
\end{figure}

This discriminant has two real roots in $\chi$, the positive (negative) root is denoted by $\chi_{+}(\mu)$ ($\chi_{-}(\mu)$). If
$\chi_{-}(\mu)<\chi <\chi_{+}(\mu)$, the equation \eref{eq26} has only one real solution that we denote by $s_{2}(\chi,\mu)$.
However, for $ \chi=\chi_{\pm}(\mu)$ we have a new equilibrium point, which undergoes a bifurcation for $\chi>\chi_{+}(\mu)$ and
$\chi<\chi_{-}(\mu)$, giving rise to the solutions denoted by $s_{3}(\chi,\mu)$ and $s_{4}(\chi,\mu)$. Figure \ref{fig1} shows
the number of real solutions of equation \eref{eq25} as a function of the collision parameters. Note that for
$\chi=\frac{9}{4}+\mu$ we have only three real solutions to \eref{eq25}, because on this curve the solutions $s_{1}$ and
$s_{3}(\chi,\mu)$ become degenerate.

By the linearization of the equations of motion \eref{eq21}, we can analyze the stability of the dynamics in the vicinity of each
equilibrium point. Although the position of the equilibrium point $s_{1}$ does not depend on the collision parameters, we can
show the following condition of stability for this equilibrium point:

\begin{equation}
  \lambda^{2}_{s_{1}}=\frac{|\Omega|^{2}}{3}(3+4\mu)(4\chi-9-4\mu)<0.
  \label{eq1ad}
\end{equation}

Therefore, considering $\mu>-\frac{3}{4}$, $s_{1}$ is stable for $\chi<\frac{9}{4}+\mu$. Note that this critical curve of
stability change coincides with the curve of degeneracy of $s_{1}$ and $s_{3}(\chi,\mu)$.

\begin{figure}[htbp]
  \centering
  \includegraphics[width=\textwidth]{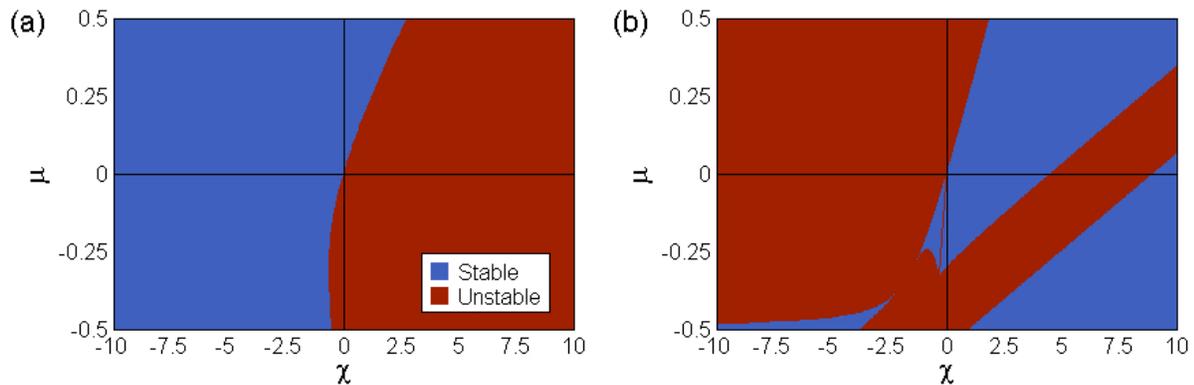}
  \caption{[colour] $\mathrm{(a)}$ Stability analysis of the equilibrium point $s_{2}(\chi,\mu)$
  as a function of collision parameters. $\mathrm{(b)}$ Stability analysis of the single depleted well states.
  Note that these are unstable equilibrium points on the curve $\chi=\mu$.}
  \label{fig2}
\end{figure}

The behaviour of the fixed point $s_{2}(\chi,\mu)$ as a function of the collision parameters is shown in figure
\ref{fig2}.$\mathrm{(a)}$. Therefore, for $|\chi|\gg|\mu|$, $s_{2}$ is generally stable (unstable) for $\chi<0$ ($\chi>0$). Also
for $|\chi|\gg|\mu|$, we can demonstrate that the equilibrium point $s_{3}$ ($s_{4}$) is unstable (stable) in its region of
existence in the parameter space, thus characterizing a \textit{saddle-node} bifurcation\footnote{We do not show the algebraic
conditions of stability for $s_{2}$, $s_{3}$ and $s_{4}$ because they are exceedingly long and complicated, due mostly to the
position dependence of these equilibrium points on $\chi$ and $\mu$.}.

\subsubsection{Single Depleted Well States}

Returning to the equations of motion, we can find two more real equilibrium points with coordinates independent of the collision
parameters:

\begin{equation}
  w_{j}=-1,\qquad w_{k}=0;
  \label{eq28}
\end{equation}

\noindent for $j,k=1,2$ and $j\neq k$. The fixed points above are known as \textit{single depleted well states}, and according to
the transformation \eref{eq22} they always possess one completely empty local mode, while the two other modes have opposite
phases and the same average population\footnote{The rotational symmetry of the trapping potential indicates the presence of a
third single depleted well state, corresponding to the complete depletion of the third well. However, according to the
transformations \eref{eq22}, this third equilibrium point would receive the coordinates $w_{1},w_{2}\rightarrow\infty$.}. The
single depleted well states have the following stability conditions, also shown in figure \ref{fig2}.$\mathrm{(b)}$:

\begin{equation}
  \fl\eqalign{
  \lambda^{2}_{SDW,\pm}=&\frac{|\Omega|^{2}}{2}
  \left\{-[9+4\mu(10+11\mu)+\chi(2+\chi)]\pm\left[(9+16\mu-\chi)\times\right.\right.\cr
  &\left.\left.\left(9+128\mu^{3}-8\mu^{2}(\chi-19)
  +\chi(5+3\chi-\chi^{2})+16\mu(4\chi+\chi^{2})\right)\right]^{\frac{1}{2}}\right\}<0.}
  \label{eq29}
\end{equation}

\subsubsection{Vortex States}

Within the classical approximation, the angular momentum of the condensate along the symmetry axis of the trapping potential is
proportional to:

\begin{equation}
  \bra N;w_{1},w_{2}|J_{S}|N;w_{1},w_{2}\ket=\frac{2N}{|w_{1}|^{2}+|w_{2}|^{2}+1}\Im\left(w_{1}-w_{2}+w\cg_{1}w_{2}\right).
  \label{eq30}
\end{equation}

Therefore, the classical angular momentum is directly related to the imaginary parts of the variables $w_{1}$ and $w_{2}$. Since
all equilibrium points found previously have real coordinates, they represent irrotational condensate configurations. However,
the equations \eref{eq21} have a further pair of fixed points:

\begin{equation}
  w_{1}=\rme^{\pm\rmi\frac{2\pi}{3}},\qquad w_{2}=\rme^{\mp\rmi\frac{2\pi}{3}}.
  \label{eq31}
\end{equation}

According to \eref{eq22}, these latter equilibrium points have the same mean occupation in all three local modes, since
$|w_{1}|=|w_{2}|=1$, but the phase difference between each pair of local condensates is $\pm\frac{2\pi}{3}$, which is the angle
of rotational symmetry of the trapping potential. The states corresponding to \eref{eq31} are known as vortex states, due to
their nonzero angular momentum, which is proportional to the total number of condensate bosons:

\begin{equation}
  \bra N;w_{1}=w_{2}\cg=\rme^{\pm\rmi\frac{2\pi}{3}}|J_{S}
  |N;w_{1}=w_{2}\cg=\rme^{\pm\rmi\frac{2\pi}{3}}\ket=\pm\sqrt{3}N.
  \label{eq32}
\end{equation}

Note that the two vortex states are equivalent, because they differ only in their sense of rotation. The stability condition for
the vortex states is given by:

\begin{equation}
  \eqalign{
  \lambda^{2}_{V,\pm}=&\frac{|\Omega|^{2}}{6}\left[
  -104\mu^{2}-16\mu(6+\chi)-3(9+4\chi)\right.\cr
  &\left.\pm\sqrt{3(3+4\mu)(3+8\mu)^{2}(9+4\mu+8\chi)}\right]<0.}
  \label{eq33}
\end{equation}

Therefore, considering $\mu>-\frac{3}{4}$, the vortex states are stable for $\chi\ge-(4\mu+9)/8$, except on the curve
$\chi=2\mu(6+11\mu)/(3+4\mu)$.

\section{Condensate Dynamics}
\label{dynamics}

\subsection{Twin-condensate Dynamics}

The integrable sub-regime of twin-condensates is a direct consequence of the symmetry of $\hat{H}$ under the permutation of
indices of the three local modes. The equivalence of the three local modes remains in the Hamiltonian \eref{eq19}, which has the
permutation symmetry of the quantities $\alpha w_{1}$, $\alpha w_{2}$ and $\alpha$, where
$\alpha=\left(|w_{1}|^{2}+|w_{2}|^{2}+1\right)^{-\frac{1}{2}}$.

Without loss of generality, we study the twin-condensate sub-regime under the restriction $w_{1}=w_{2}$, because the invariant
surfaces in \eref{eq24} are dynamically equivalent. Applying this condition to the coherent states in \eref{eq17}, we obtain:

\begin{equation}
  \eqalign{
  |N;w_{1}=w_{2}\ket&=\frac{1}{\sqrt{N!}}\left[\frac{w_{1}a\dg_{1}+w_{1}a\dg_{2}+a\dg_{3}}
  {\sqrt{2|w_{1}|^{2}+1}}\right]^{N}|0\ket\cr
  &=\suml_{m_{1}+m_{2}=N}\frac{\sqrt{N!}}{m_{1}!m_{2}!}
  \frac{(w_{1}a\dg_{1}+w_{1}a\dg_{2})^{m_{1}}(a\dg_{3})^{m_{2}}}{(2|w_{1}|^{2}+1)^{\frac{N}{2}}}|0\ket.}
  \label{eq34}
\end{equation}

Now it is convenient to change the basis of the single-particle Hilbert space. Such transformation may be described by the
following linear combinations of the bosonic operators:

\begin{equation}
  \left\{\eqalign{
  b\dg_{1}&=\frac{a\dg_{1}+a\dg_{2}}{\sqrt{2}}\cr
  b\dg_{2}&=a\dg_{3}\cr
  b\dg_{3}&=\frac{a\dg_{1}-a\dg_{2}}{\sqrt{2}}}
  \right.
  \label{eq35}
\end{equation}

The operator $b\dg_{1}$ ($b\dg_{3}$) creates a particle in the equiprobable superposition of the local states $|u_{1}\ket$ and
$|u_{2}\ket$ arranged with identical (opposite) phases. So $b\dg_{1}$ is responsible for the occupation of the twin-condensates,
while $b\dg_{2}$ represents the \textit{solitary mode}, which is characterized by the single-particle state $|u_{3}\ket$. Note
that $b\dg_{3}$ is also the creation operator in the state $|e_{3}\ket$, found in \eref{eq3}. Applying the transformation
\eref{eq35} in \eref{eq34}, we have:

\begin{equation}
  \eqalign{
  |N;w_{1}=w_{2}\ket&=\suml_{m_{1}+m_{2}=N}\left(\frac{N!}{m_{1}!m_{2}!}\right)^{\frac{1}{2}}
  \frac{(\sqrt{2}w_{1})^{m_{1}}|m_{1},m_{2},0\ket}{(2|w_{1}|^{2}+1)^{\frac{N}{2}}}\cr
  &=\frac{1}{\sqrt{N!}}\left[\frac{(\sqrt{2}w_{1})b\dg_{1}+b\dg_{2}}
  {\sqrt{2|w_{1}|^{2}+1}}\right]^{N}|0\ket\;=\;|N;\sqrt{2}w_{1}\ket_{\mathrm{su}(2)}.}
  \label{eq36}
\end{equation}

The set of states $\{| m_{1},m_{2},m_{3}\ket\}$ constitutes a new basis for the bosonic Fock space, where $m_{j}$ represents the
eigenvalue of the number operator $b\dg_{j}b_{j}$. Note that the $\mathrm{SU}(3)$ coherent state restricted to the subspace
twin-condensates has no occupation in the mode related to the operator $b\dg_{3}b_{3}$, also called the \textit{opposite phase
mode}, evidencing the phase equality between the twin-condensates in the classical dynamics. Besides, the coherent state
\eref{eq36} is also a $\mathrm{SU}(2)$ coherent state with complex coordinate
$\sqrt{2}w_{1}=\rme^{-\rmi\phi}\tan\frac{\theta}{2}$, indicating that the invariant surface of twin-condensates is topologically
isomorphic to the unit sphere ($S^{2}$) \cite{su2coher}.

The states with zero mean occupation in the opposite phase mode, given by $\bra\psi|b\dg_{3}b_{3}|\psi\ket=0$ or
$|\psi\ket=\suml_{m_{1}+m_{2}=N}c_{m_{1},m_{2}}|m_{1},m_{2},0\ket$ with $c_{m_{1},m_{2}}\in\mathds{C}$, satisfy the following
identities:

\begin{equation}
  \eqalign{
  \bra\psi| a\dg_{1}a_{1}+a\dg_{2}a_{2}|\psi\ket=\bra\psi|a\dg_{1}a_{2}+a\dg_{2}a_{1}|\psi\ket;\cr
  \bra\psi| a\dg_{1}a_{1}|\psi\ket=\bra\psi|a\dg_{2}a_{2}|\psi\ket;\cr
  \frac{d}{dt}\bra\psi|Q_{1}|\psi\ket=0.}
  \label{eq37}
\end{equation}

Therefore, the quantum dynamics preserves the population equality between the twin-condensates, because the average of the
population imbalance operator $Q_{1}$ remains zero during the quantum evolution for all the initial states in the subspace
generated by the basis $\{|m_{1},m_{2},0\ket\}$. However, the quantum dynamics does not preserve the identical phases of
twin-condensates, unlike the classical dynamics, because the average of $b\dg_{3}b_{3}$ does not remain constant, in spite of its
initial zero value.

At this point it is interesting to define the generators of another subalgebra of $\mathrm{su}(3)$ isomorphic to
$\mathrm{su}(2)$, but related to the modes present in the twin-condensate sub-regime:

\begin{equation}
  S_{x}=\frac{b\dg_{1}b_{2}+b\dg_{2}b_{1}}{2},\qquad S_{y}=\rmi\frac{b\dg_{2}b_{1}-b\dg_{1}b_{2}}{2},
  \qquad S_{z}=\frac{b\dg_{1}b_{1}-b\dg_{2}b_{2}}{2}.
  \label{eq38}
\end{equation}

Now, employing the identities \eref{eq37}, we can show that:

\begin{equation}
  \bra\psi|S_{z}|\psi\ket=\frac{1}{2}\bra\psi|a\dg_{1}a_{1}+a\dg_{2}a_{2}-a\dg_{3}a_{3}|\psi\ket.
  \label{eq39}
\end{equation}

Thus, $S_{z}$ is the population imbalance operator between the twin-condensates and the solitary mode, but only if the average
occupation of the opposite phase mode is zero.

\begin{figure}[htbp]
  \centering
  \includegraphics[width=\textwidth]{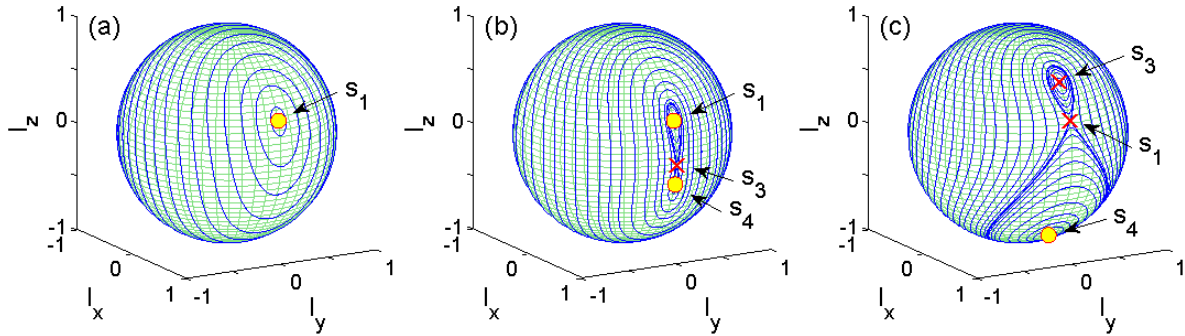}
  \caption{[colour] Classical trajectories in the invariant subspace of twin-condensates
  $w_{1}=w_{2}=\frac{1}{\sqrt{2}}\rme^{-\rmi\phi}\tan\frac{\theta}{2}$ for $N=30$, $|\Omega|=1$,
  $\mu=0$ and self-collision parameter values: $\mathrm{(a)}$ $\chi=1.5$, $\mathrm{(b)}$ $\chi=1.98$ and
  $\mathrm{(c)}$ $\chi=3$. The [yellow] circles ([red] crosses) indicate the location of stable (unstable) fixed points
  with respect to the complete four-dimensional phase space of the three-mode condensate. The angles $\phi$ and $\theta$
  are similar to the usual spherical coordinate angles, except that, for convenience, we take the origin of $\theta$ along
  the \textit{negative} z-axis.} \label{fig3}
\end{figure}

Figure \ref{fig3} displays the classical dynamics of the three-mode model in the invariant subspace of twin-condensates, which is
represented by the unit sphere, for various values of the self-collision rate $\chi$, but neglecting the presence of
cross-collisions. Observe that the Cartesian coordinates directly represent the normalized classical averages of the
$\mathrm{SU}(2)$ generators:

\begin{equation}
  \left\{\eqalign{
  I_{x}=\frac{2}{N}\bra N;w_{1}=w_{2}|S_{x}|N;w_{1}=w_{2}\ket=\sin\theta\cos\phi\cr
  I_{y}=\frac{2}{N}\bra N;w_{1}=w_{2}|S_{y}|N;w_{1}=w_{2}\ket=\sin\theta\sin\phi\cr
  I_{z}=\frac{2}{N}\bra N;w_{1}=w_{2}|S_{z}|N;w_{1}=w_{2}\ket=-\cos\theta
  }\right.
  \label{eq40}
\end{equation}

Figure \ref{fig3}.$\mathrm{(a)}$, for $\chi=1.5<\chi_{+}(\mu=0)\approx1.971$, shows the classical dynamics in the absence of the
equilibrium points $s_{3}$ and $s_{4}$. Note that all orbits surround the stable equilibrium point $s_{1}$ and, therefore, the
population imbalance varies around the value $I_{z}(w_{1}=w_{2}=1)=\frac{1}{3}$, which represents an identical average population
in all three local modes. Thus, there is no preferential occupation of any local mode, and this behaviour characterizes the
dynamical regime known as \textit{Josephson oscillation} (JO). In figure \ref{fig3}.$\mathrm{(b)}$ we show the classical dynamics
for $\chi=1.98$, i.e., a self-collision rate value slightly higher than the critical bifurcation parameter. The bifurcation is
accompanied by the appearance of a \textit{separatrix}, which crosses the unstable equilibrium point $s_{3}$. The separatrix is
the boundary between the orbits of the JO regime and the trajectories around the new stable equilibrium point $s_{4}$. The
trajectories near $s_{4}$ vary around negative values of $I_{z}$, indicating a preferential occupation of the third well.
Therefore, the tunneling between twin-condensates and solitary mode is suppressed, giving rise to the effect known as
\textit{macroscopic self-trapping} (MST). The JO and MST sub-regimes are also present in the double-well model, whose classical
approximation with $\mathrm{SU}(2)$ coherent states is quite similar to the dynamics of twin-condensates \cite{2well}.

Figure \ref{fig3}.$\mathrm{(c)}$ displays the dynamics of twin-condensates for $\chi=3$, a parameter value significantly higher
than $\chi_{+}$. For $\chi>\frac{9}{4}$ the fixed point $s_{1}$ becomes unstable and takes the place of $s_{3}$ in the
separatrix. Although the orbits around $s_{3}$ in the subspace of twin-condensates are regular, we should note that this
equilibrium point is not a stable center in the complete phase space, since $s_{3}$ behaves like a saddle point in the directions
orthogonal to the surface $w_{1}=w_{2}$. Therefore, the self-trapping dynamics which favors the occupation of the
twin-condensates, characterized by the regular trajectories with $I_{z}>\frac{1}{3}$ in the vicinity of $s_{3}$, is restricted to
the invariant surface. Besides, we observe that an increase in self-collision rate is accompanied by an expansion of the region
on $S^{2}$ occupied by MST orbits.

\begin{figure}[hbtp]
  \centering
  \includegraphics[width=0.8\textwidth]{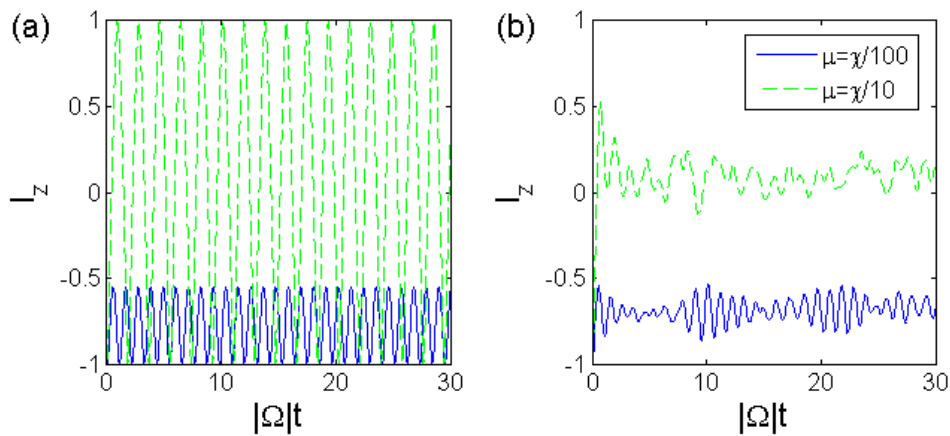}
  \caption{[colour] Left (right) figure shows the classical (quantum) time evolution of the population imbalance
  $I_{z}$ for the initial state $|N,w_{1}=w_{2}=0\ket=|0,0,N\ket$. The solid [blue] (dashed [green]) curve
  displays the results for $N=30$, $|\Omega|=1$, $\chi=4$ and $\mu=\frac{\chi}{100}$ ($\mu=\frac{\chi}{10}$).}
  \label{fig4}
\end{figure}

Figure \ref{fig4} shows the effect of cross-collisions on the dynamics of the mean population imbalance $I_{z}$ for the initial
coherent state $|N,w_{1}=w_{2}=0\ket=|0,0,N\ket$, corresponding to the south pole in figure \ref{fig3}. The solid [blue] (dashed
[green]) curve in figure \ref{fig4}.$\mathrm{(a)}$ displays the classical time evolution of $I_{z}$ for $N=30$, $|\Omega|=1$,
$\chi=4$ and $\mu=\frac{\chi}{100}$ ($\mu=\frac{\chi}{10}$). Notice that the greater value of the cross-collision parameter
inhibits the MST, due to its correspondent increase of the effective tunneling rate $\Omega'=\Omega(1+2\mu)$. Therefore, the
condensate dynamics can be completely changed by the cross-collisions, even when the collision parameters $\mu$ and $\chi$ have
different magnitudes.

Figure \ref{fig4}.$\mathrm{(b)}$ exhibits the quantum time evolution of $I_{z}$ for a direct comparison with the classical
results. Note that, unlike previous classical results, the amplitude of quantum oscillations does not remain constant during the
evolution of the condensate, evidencing a quantitative disagreement between the two approaches. However, the solid [blue] (dashed
[green]) curve properly characterizes the MST (JO) regime, because the fluctuations in the population imbalance remain around
$I_{z}<0$ ($I_{z}=\frac{1}{3}$). Consequently, the qualitative agreement between the classical and quantum approaches persists,
even for a number of particles as small as $N=30$.

\subsubsection{Generalized Purity associated with the $\mathrm{su}(3)$ algebra}

The classical equations of motion \eref{eq21} do not depend on the total number of trapped bosons, since they represent the
dynamics in the macroscopic limit $N\rightarrow\infty$. Therefore, the qualitative agreement between the two approaches would not
be expected for a small number of particles such as $N=30$, unlike observed in figure \ref{fig4}. Although mean field and
classical theories are quite common in the treatment of condensate dynamics, little is known about the quality of these
approximations with respect to the exact quantum calculation for a microscopic or mesoscopic condensate. The classical dynamics
is almost exact for a macroscopic number of particles, but for a mesoscopic or microscopic number of bosons we must be able to
evaluate \textit{quantitatively} the quality of our approximations as a function of $N$ and the propagation time, because in an
usual experiment the number of condensate particles can range from a few hundred to the order of $10^{10}$ bosons
\cite{condensate}.

Our classical approximation consists of restricting the system time evolution to the nonlinear subspace of coherent states.
However, the exact quantum dynamics for few particles can promote the departure of an initial coherent state from the classical
subspace, thus introducing quantitative errors in our approximation. The generalized purity associated with the $\mathrm{su}(3)$
algebra, which is a measure capable of quantifying the proximity of a given $N$-particle state $|\psi\ket$ to the classical
subspace, is given by \cite{purity}:

\begin{equation}
  \fl \mathcal{P}_{\mathrm{su}(3)}(|\psi\ket)=\frac{9}{N^{2}}\left(\frac{\bra\psi|Q_{1}|\psi\ket^{2}}{3}
  +\frac{\bra\psi|Q_{2}|\psi\ket^{2}}{4}
  +\suml_{j=1}^{3}\frac{\bra\psi|P_{j}|\psi\ket^{2}}{12}
  +\suml_{j=1}^{3}\frac{\bra\psi|J_{j}|\psi\ket^{2}}{12}\right).
  \label{eq23}
\end{equation}

The generalized purity is limited to the interval $0\leq\mathcal{P}_{\mathrm{su}(3)}(|\psi\ket)\leq1$, but we have
$\mathcal{P}_{\mathrm{su}(3)}(|\psi\ket)=1$ if and only if $|\psi\ket$ is a coherent state. On the other hand, the value of
$\mathcal{P}_{\mathrm{su}(3)}(|\psi\ket)$ is decreasing with the ``distance'' of $|\psi\ket$ to the subspace of coherent states.
Therefore, $\mathcal{P}_{\mathrm{su}(3)}(|\psi\ket)$ is a \textit{classicality} measure, because the purity value is increasing
with the classical character of $|\psi\ket$, taking the coherent states as the most classical pure states, due to their minimum
uncertainty on the phase space.

\begin{figure}[htbp]
  \centering
  \includegraphics[width=0.8\textwidth]{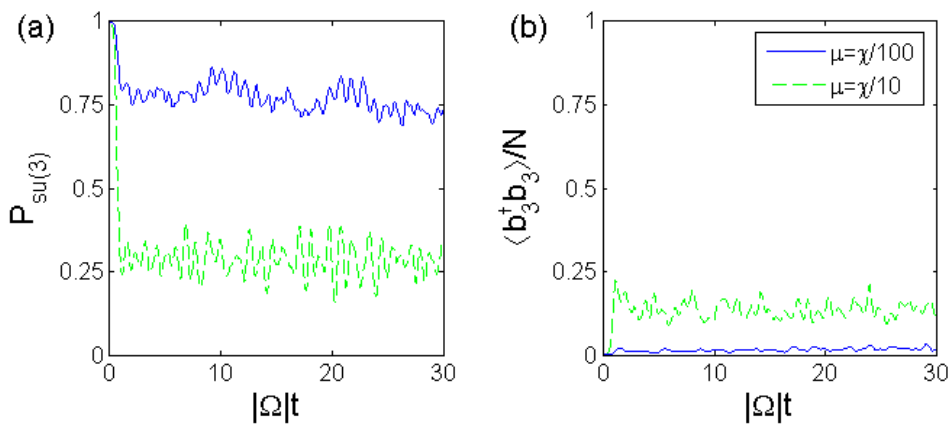}
  \caption{[colour] Left (right) figure shows the time evolution of $\mathcal{P}_{\mathrm{su}(3)}$ ($\bra b\dg_{3}b_{3}\ket$)
  for an immediate comparison with the figure \ref{fig4}. The solid [blue] (dashed [green]) curve shows the results for
  $N=30$, $|\Omega|=1$, $\chi=4$, $\mu=\frac{\chi}{100}$ ($\mu=\frac{\chi}{10}$) and initial state $|N,w_{1}=w_{2}=0\ket=|0,0,N\ket$.}
  \label{fig5}
\end{figure}

Figure \ref{fig5}.$\mathrm{(a)}$ displays the evolution of $\mathcal{P}_{\mathrm{su}(3)}$ for the same initial state and
parameters of figure \ref{fig4}. Note that the state initially suffers a sudden purity loss, which is responsible for the
quantitative disagreement between the classical and quantum results. Therefore, only for a very short time interval the classical
approximation roughly coincides with the quantum dynamics, when considering a small number of particles. However, after a brief
initial period of purity loss, $\mathcal{P}_{\mathrm{su}(3)}$ only exhibits fluctuations around a stable value. The purity loss
in the MST regime is inferior to the JO regime, indicating the better quality of the classical approximation in the self-trapping
dynamics. The orbits associated with the JO regime traverse a larger region of the phase space compared with the localized MST
trajectories, as shown in figure \ref{fig3}. Therefore, the quantum states related to the JO regime have greater delocalization
(uncertainty) on the phase space, which is responsible for their accentuated classicality loss.

Within the classical approximation, the invariant subspaces of twin-condensates are defined by the equality of phase and mean
occupation between two local modes. Although the quantum dynamics preserves the population equality between identical local
condensates, according to the equation \eref{eq37}, the phase difference does not remain zero. In general, using the
transformation \eref{eq35} in the Hamiltonian \eref{eq8}, we can show that:

\begin{equation}
  \frac{d}{dt}\bra N;w_{1}=w_{2}|b\dg_{3}b_{3}|N;w_{1}=w_{2}\ket\neq0. \label{eq41}
\end{equation}

Except for $\chi=\mu=0$ or $N\rightarrow\infty$, when the classical approximation is exact. Therefore, the subspaces with no
phase difference between two local modes are not quantum invariant. The population dynamics of the opposite phase mode for the
states $|N,w_{1}=w_{2}\ket$ can not be described classically, and therefore it represents a process of purity (classicality)
loss. Figure \ref{fig5}.$\mathrm{(b)}$ shows the behaviour of $\bra b\dg_{3}b_{3}\ket$ for the same initial state and parameters
of the previous figure. In comparison with the figure \ref{fig5}.$\mathrm{(a)}$, we observe that a higher occupation of the
opposite phase mode is accompanied by a greater purity loss. Therefore, the nonzero occupation of the opposite phase mode is
partially responsible for breaking the quantum-classical correspondence in the sub-regime of twin-condensates. Moreover, the
quantity $\bra b\dg_{3}b_{3}\ket$ can be used as a measure of validity (quality) of the classical approximation in the invariant
subspaces.

\subsection{Single Depleted Well Dynamics}

The sub-regime of twin-condensates is very similar to the dynamics of a condensate in a double-well potential, since both models
have several features in common, such as the integrability, the $\mathrm{SU}(2)$ coherent states, and especially the dynamical
transition from JO to MST. Therefore, the single depleted well states (SDWS) represent the first example of an entirely new
population dynamics for the three-mode model.

According to the transformation \eref{eq22}, the SDWS have a completely empty local mode, while the two other modes remain with
the same mean occupation and opposite phases. Similarly to the self-trapping regime, the SDWS represent a dynamical effect of
tunneling suppression, but the opposite phases between the preferentially occupied local modes distinguish the latter regime.

Due to the rotational symmetry of the trapping potential, all the SDWS are equivalent. Thus, without loss of generality, we
restrict our results to the region of phase space in the vicinity of $(w_{1}=-1,w_{2}=0)$, where the second local mode is empty.

\begin{figure}[htbp]
  \centering
  \includegraphics[width=0.8\textwidth]{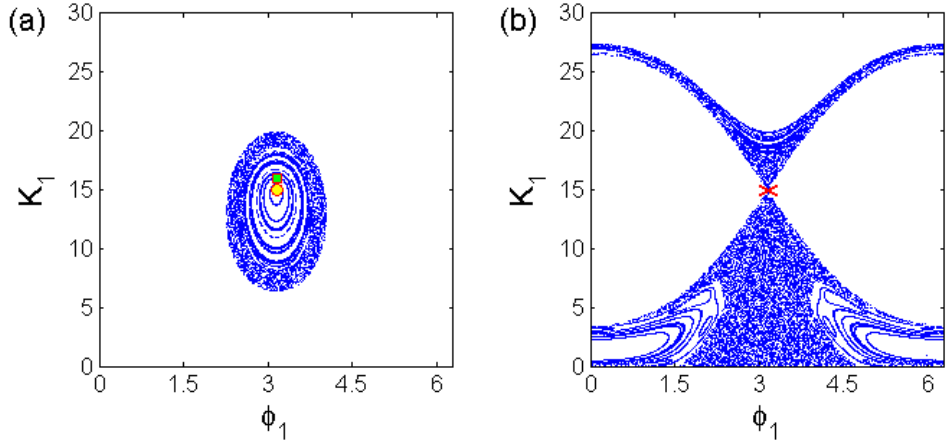}
  \caption{[colour] Left (right) figure shows the Poincar\'e section at $\phi_{2}=0$ for $N=30$,
  $|\Omega|=1$, $\chi=5$ ($\chi=-5$), $\mu=\frac{\chi}{100}$, considering only those trajectories with the same
  energy of the fixed point $(w_{1}=-1,w_{2}=0)$. The [yellow] circle ([red] cross) shows the position of the
  stable (unstable) equilibrium point in the section.}
  \label{fig6}
\end{figure}

The SDWS have fixed positions in the phase space, but their stability varies as a complicated function of the collision
parameters $\chi$ and $\mu$, according to the condition \eref{eq29} and the figure \ref{fig2}.$\mathrm{(b)}$. Figure
\ref{fig6}.$\mathrm{(a)}$ shows the Poincar\'e section at $\phi_{2}=0$ for $\chi=5$ and $\mu=\frac{\chi}{100}$, considering only
the trajectories with the same energy of the SDWS. Note that the section presents the stable SDWS located in the center of a
bundle of regular orbits. The energetically accessible region of phase space is filled by orbits with a behaviour similar to the
SDWS, since they exhibit only small fluctuations in the average occupation of the second local mode, while the other two modes
oscillate in opposite phase around $\frac{N}{2}$.

Figure \ref{fig6}.$\mathrm{(b)}$ shows the Poincar\'e section in the vicinity of the SDWS for $\chi=-5$ and
$\mu=\frac{\chi}{100}$. Here we see an unstable SDWS inserted in a chaotic region of phase space. We also observe that the SDWS
is the boundary between two distinct sets of chaotic orbits, because the trajectories above (below) the equilibrium point in this
section preserve the condition $K_{1}>\frac{N}{2}$ ($K_{1}<\frac{N}{2}$) during their time evolution. Unlike the stable case, the
energetically accessible region of phase space in the vicinity of the fixed point does not show a behaviour similar to the SDWS,
since there is no persistent depletion of the second local mode.

\begin{figure}[htbp]
  \centering
  \includegraphics[width=0.8\textwidth]{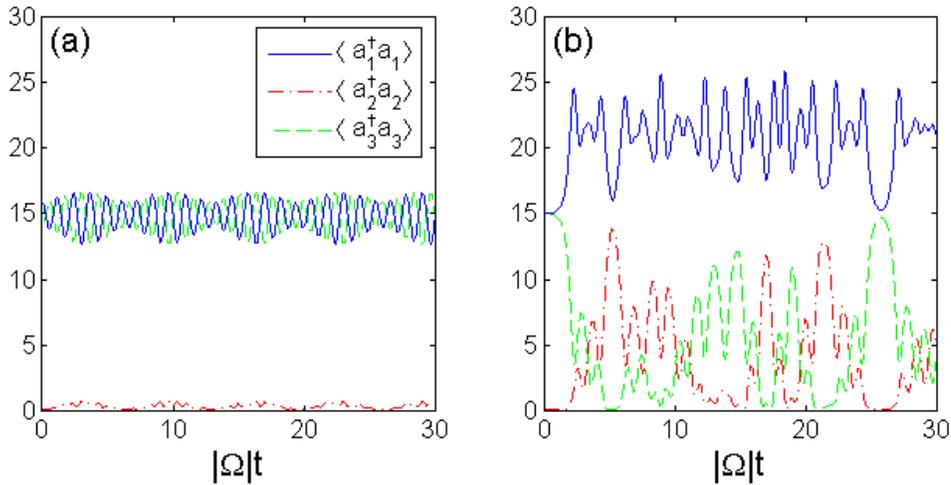}
  \caption{[colour] Left (right) figure shows the classical population dynamics for an initial condition
  near one of the SDWS, considering $N=30$, $|\Omega|=1$, $\chi=5$  ($\chi=-5$), $\mu=\frac{\chi}{100}$.}
  \label{fig7}
\end{figure}

The regular population dynamics in the vicinity of the SDWS for $\chi=5$ and $\mu=\frac{\chi}{100}$ is illustrated in figure
\ref{fig7}.$\mathrm{(a)}$, employing the initial condition indicated by the square [green] marker in figure
\ref{fig6}.$\mathrm{(a)}$. As expected, the second local mode remains almost empty, while the two remaining modes exhibit
periodic population inversions around $\frac{N}{2}$. Figure \ref{fig7}.$\mathrm{(b)}$ shows the chaotic population dynamics for
$\chi=-5$ and $\mu=\frac{\chi}{100}$, given an initial condition very close to the SDWS with $K_{1}>\frac{N}{2}$. Note that the
second local mode does not remain empty, and it presents successive population inversions with the third mode, while the first
mode holds more than half of the trapped bosons.

\begin{figure}[htbp]
  \centering
  \includegraphics[width=\textwidth]{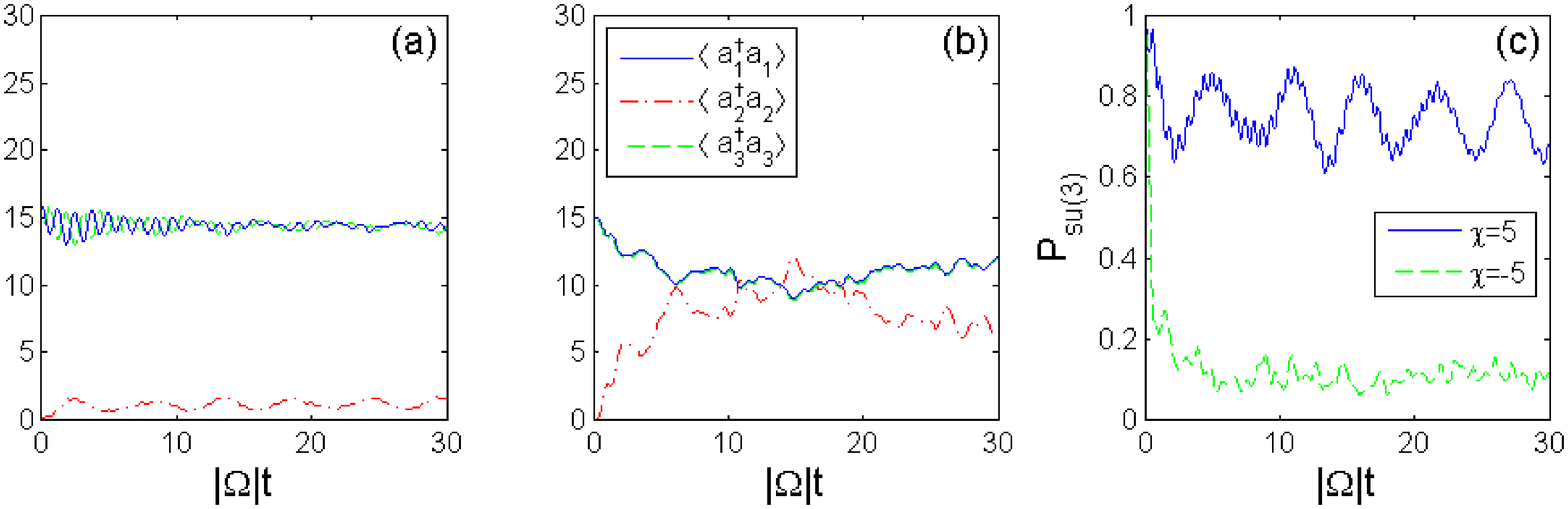}
  \caption{[colour] Left (center) figure shows the quantum evolution
  of the average occupations in the three local modes for $N=30$, $|\Omega|=1$, $\chi=5$ ($\chi=-5$), $\mu=\frac{\chi}{100}$,
  considering as the initial state the coherent state centered at the initial condition of the trajectory in
  figure \ref{fig7}.$\mathrm{(a)}$ (figure \ref{fig7}.$\mathrm{(b)}$). Notice in $\mathrm{(b)}$ that the curves for
  $\bra a\dg_{1}a_{1}\ket$ and $\bra a\dg_{3}a_{3}\ket$ are almost superimposed. The solid [blue] and dashed [green] curves
  in $\mathrm{(c)}$ represent the dynamics of $\mathcal{P}_{\mathrm{su}(3)}$ corresponding to the results
  displayed in $\mathrm{(a)}$ and $\mathrm{(b)}$, respectively.}
  \label{fig8}
\end{figure}

For direct comparison with the previous classical results, the exact quantum population dynamics is shown in figure \ref{fig8},
where we used as initial states the coherent states centered at the initial conditions of figure \ref{fig7}. Figure
\ref{fig8}.$\mathrm{(a)}$ displays the quantum population dynamics corresponding to the regular trajectory near the SDWS. The
mean occupation in the second local mode is slightly higher than in the classical approximation, but the preferential occupation
in the other two modes still prevails. Therefore, there is a fair agreement between the exact quantum result and the classical
approximation when we consider a regular orbit close to the stable SDWS. However, the quantum dynamics corresponding to the
chaotic trajectory, shown in figure \ref{fig8}.$\mathrm{(b)}$, has little resemblance to its classical analogue. Unlike the
classical approximation, there is no preferential occupation of the first mode, which displays values below $\frac{N}{2}$. On the
other hand, the second well does not remain empty, as expected for the irregular dynamics near the SDWS.

The dynamics of $\mathcal{P}_{\mathrm{su}(3)}$ confirms that the classical approximation is more accurate for the regular
trajectory, as we see in figure \ref{fig8}.$\mathrm{(c)}$. The state associated with the chaotic dynamics has a large purity
loss, which corresponds to a large delocalization in the phase space. Conversely, the time evolution of the state under the
regular dynamics preserves its similarity to the coherent states, i.e., the state remains well localized in the phase space and
exhibits relatively high values of purity during its evolution.

\subsection{Vortex Dynamics}

The vortex states are the only equilibrium points of the classical approximation with nonzero imaginary parts. Thus, they are the
only ones with nonzero angular momentum along the symmetry axis of the trapping potential. To emphasize the imaginary parts of
the complex variables $w_{j}$ during their time evolution, we introduce a new set of canonical variables:

\begin{equation}
w_{j}=\frac{q_{j}+\rmi p_{j}}{\sqrt{2N-q_{1}^{2}-p_{1}^{2}-q_{2}^{2}-p_{2}^{2}}}. \label{eq42}
\end{equation}

The real variables $q_{j}$ and $p_{j}$ are directly related to the real and imaginary parts of $w_{j}$, respectively. Note also
that these new dynamical variables must satisfy the condition $2N\geq q_{1}^{2}+p_{1}^{2}+q_{2}^{2}+p_{2}^{2}$. We show below
only the results for the vortex state parametrized by $w_{1}=\rme^{\rmi\frac{2\pi}{3}}=w_{2}\cg$, since the two vortex states
differ only by the sense of rotation, as previously discussed.

\begin{figure}[htbp]
  \centering
  \includegraphics[width=0.8\textwidth]{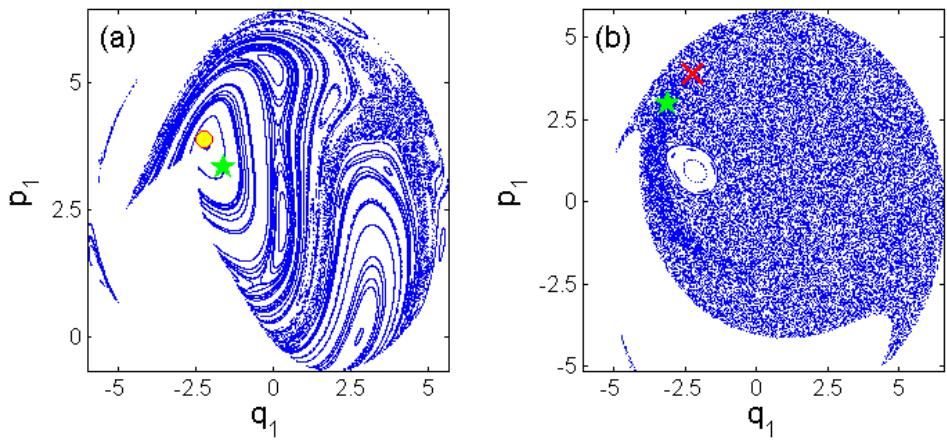}
  \caption{[colour] Left (right) figure shows the Poincar\'e section at $p_{2}=-\sqrt{N/2}$ for
  $N=30$, $|\Omega|=1$, $\chi=-1$ ($\chi=-5$) and $\mu=\frac{\chi}{100}$, considering only those trajectories
  with the same energy of the vortex state $w_{1}=e^{i\frac{2\pi}{3}}=w_{2}\cg$. The stable (unstable) equilibrium
  point is represented by the [yellow] circle ([red] cross).}
  \label{fig9}
\end{figure}

Figure \ref{fig9} shows the dynamics in the vicinity of the vortex state for two completely different situations. For $\chi=-1$
and $\mu=\frac{\chi}{100}$ the Poincar\'e section at $p_{2}=-\sqrt{N/2}$ displays the stable vortex state in a regular region of
phase space. However, for $\chi=-5$ and $\mu=\frac{\chi}{100}$, the vortex state is unstable and the dynamics of the system is
almost completely chaotic.

\begin{figure}[htbp]
  \centering
  \includegraphics[width=\textwidth]{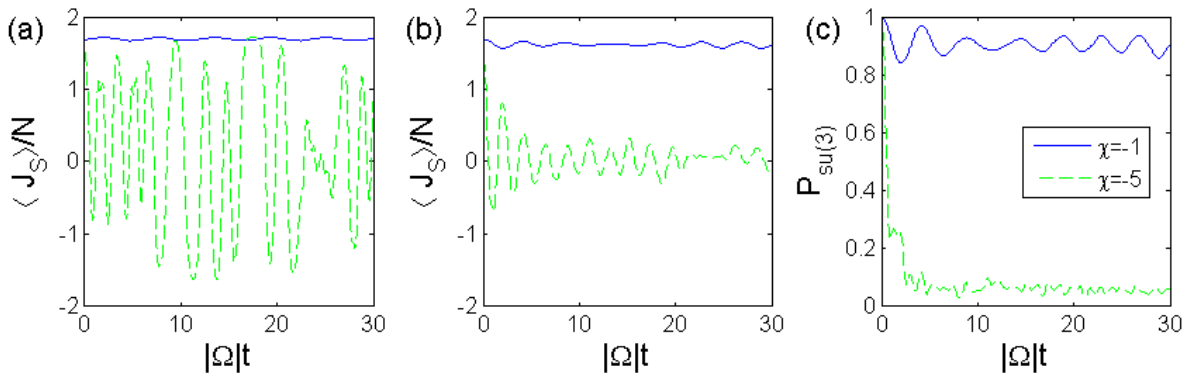}
  \caption{[colour] Left (middle) figure exhibits the classical (quantum) dynamics of $\bra J_{S}\ket$,
  quantity directly proportional to the angular momentum of the condensate along the symmetry axis of the
  trapping potential. The solid [blue] (dashed [green]) curve shows the time evolution for $N=30$, $|\Omega|=1$,
  $\chi=-1$ ($\chi=-5$), $\mu=\frac{\chi}{100}$ and the initial condition indicated by the
  [green] star marker in figure \ref{fig9}.$\mathrm{(a)}$ (figure \ref{fig9}.$\mathrm{(b)}$). Right figure shows the
  dynamics of $\mathcal{P}_{\mathrm{su}(3)}$ for the same initial conditions.}
  \label{fig10}
\end{figure}

According to the equation \eref{eq15}, the angular momentum of the condensate along the symmetry axis of the trapping potential
is directly proportional to the $\mathrm{su}(3)$ operator $J_{S}=J_{1}+J_{2}+J_{3}$. Therefore, the rotational dynamics of the
condensate is illustrated in Figure \ref{fig10}, where we consider the initial conditions indicated by the [green] star markers
in figure \ref{fig9}. The solid [blue] (dashed [green]) curve in figure \ref{fig10}.$\mathrm{(a)}$ shows the classical dynamics
of a regular (chaotic) trajectory near the vortex state. Note that only the regular trajectory represents a persistent collective
rotation of the condensate. Therefore, the rotation of the condensate in a preferential sense is present only when the vortex
state is stable. The chaotic trajectory does not exhibit preferential sense of rotation, but presents oscillations of angular
momentum with large amplitudes. However, the maximum value of $\bra J_{S}(t)\ket$ in both trajectories remains bounded by
$\sqrt{3}N$, which corresponds to the mean value of $J_{S}$ for the vortex state, according with \eref{eq32}.

Figure \ref{fig10}.$\mathrm(b)$ shows the quantum dynamics of rotation, where we consider the initial coherent states centered on
the initial conditions of figure \ref{fig10}.$\mathrm{(a)}$. Note that even for $N=30$, a relatively small number of trapped
bosons, the quantum results of $\bra J_{S}(t)\ket$ associated with the regular trajectory demonstrate good agreement with the
classical approximation. However, in the chaotic case, the amplitude of the oscillations are strongly attenuated. Thus, as
expected, the approximation is not quantitatively satisfactory in the chaotic regime, although the absence of a preferential
sense of rotation is evidenced in both approaches. Figure \ref{fig10}.$\mathrm{(c)}$ displays the dynamics of
$\mathcal{P}_{\mathrm{su}(3)}$ for the two orbits previously selected. Note that the purity loss in the chaotic regime is much
faster and more intense when compared to the regular system, confirming the differences found in the precision of the classical
approximation.

\section{Conclusion}
\label{conclusion}

In the present work, we have investigated the quantum dynamics of a Bose-Einstein condensate confined in a symmetric triple-well
potential employing an approximation based on the TDVP and the $\mathrm{SU}(3)$ coherent states. We have exploited maximally the
use of coherent states as initial states chosen to be centered in appropriate points of the classical manifold, particularly near
the equilibrium points of the classical equations of motion. This scheme allowed us to capture in the complete quantum
calculations some of the particularities of the classical solutions.

Moreover, in order to evaluate the quality of the classical Hamiltonian evolution, as compared to the quantum results, we
employed the purity associated with the $\mathrm{su}(3)$ algebra. The generalized purity measures the departure from the subspace
of coherent states, in which our classical approximation is restricted, thus enabling us to quantify the quality of the
approximation. The classical approximation quality depends not only on the total number of particles, but also on the location in
phase space of the initial coherent state and the regularity of the Hamiltonian flow in its vicinity. We concluded that even for
a relatively small number of particles such as $N=30$, which is very far from the classical-macroscopic limit, the classical
results obtained for the \textit{regular} dynamical regime display a fair agreement with the quantum time evolution, unlike the
\textit{chaotic} regime.

Considering the sub-regime of \textit{twin-condensates}, we have shown that the self-trapping dynamics may be suppressed by the
cross-collisions effects on the effective tunneling rate, which can become quite important for a large number of trapped bosons.
Since this sub-regime is classically integrable, its classical approximation exhibits excellent results. Whereas, the
\textit{single depleted well} regime, which is characterized by the complete depletion of a local mode, can be effectively
eliminated in phase space when its corresponding fixed points become unstable, giving way to chaos. In the chaotic regime, a
single Hamiltonian trajectory clearly does \textit{not} account for the quantum dynamics of an initial coherent state at long
times. This fact is precisely indicated by the generalized purity. In the same line of reasoning, the persistent collective
rotation in a preferential sense, presented in the vicinity of the \textit{vortex} states in phase space, can only be properly
observed when the corresponding fixed points are stable.

\section*{Acknowledgments}

T.F.V. would like to thank Marcus A. M. de Aguiar for many insightful discussions. We also acknowledge the financial support from
FAPESP under Grant No. 2008/09491-9 and CNPq under Grant No. 304041/2007-6. This work is partially supported by the Brazilian
National Institute of Science and Technology of Quantum Information (INCT-IQ).

\end{document}